\documentclass[onecolumn,trackchanges]{aastex7}

\usepackage{amsmath}
\usepackage{amssymb}
\usepackage{xcolor}
\usepackage{graphicx}
\usepackage{float}
\usepackage{float}
\usepackage{subfigure}
\usepackage{hyperref}
\usepackage{comment}
\usepackage{url}
\usepackage{makecell}   
\usepackage{array}
\usepackage{multirow}
\usepackage{array} 
\begin{document}

\title{Shear Particle Acceleration in Structured Gamma-Ray Burst Jets: IV. Thermal {\em vs.} Non-thermal Emission of the Jet Cocoon}

\correspondingauthor{Xiao-Li Huang, En-Wei Liang}

\author[0009-0001-8025-3205]{Zi-Qi Wang}
\affiliation{Guangxi Key Laboratory for Relativistic Astrophysics, School of Physical Science and Technology, Guangxi University, Nanning 530004, People’s Republic of China}
\email[show]{ziqi.wang@st.gxu.edu.cn} 

\author[0000-0002-9725-7114]{Xiao-Li Huang}
\affiliation{School of Physics and Electronic Science, Guizhou Normal University, Guiyang 550025, People’s Republic of China}
\affiliation{National Astronomical Observatories, Chinese Academy of Sciences, Beijing 100101, People’s Republic of China}
\email[show]{xiaoli.huang@gznu.edu.cn}

\author[0000-0001-6863-5369]{Hai-Ming Zhang}
\affiliation{Guangxi Key Laboratory for Relativistic Astrophysics, School of Physical Science and Technology, Guangxi University, Nanning 530004, People’s Republic of China}
\email[show]{hmzhang@gxu.edu.cn}

\author[0000-0002-7044-733X]{En-Wei Liang}
\affiliation{Guangxi Key Laboratory for Relativistic Astrophysics, School of Physical Science and Technology, Guangxi University, Nanning 530004, People’s Republic of China}
\email[show]{lew@gxu.edu.cn}

\begin{abstract}
A distinct thermal or quasi-thermal spectral component is occasionally observed in gamma-ray burst (GRB) prompt emission spectra. 
Taking GRB 090902B as a case study, we investigate its origin within a structured jet framework, in which the outflow consists of an ultra-relativistic uniform core surrounded by a structured cocoon. 
In the weak-scattering regime with inefficient shear acceleration, electrons pre-energized in the thin jet-cocoon interaction layer are further heated in the mixed jet-cocoon (MJC) region, forming a quasi-thermal electron distribution.
Parameterizing the radial temperature profile of electrons as a power law with index $q_T$, we demonstrate that both the peak flux and spectral width of the thermal component are sensitive to maximum temperature $T_{\max}$ and $q_T$. 
Combined with the synchrotron emission of shock-accelerated electrons in the jet core, our model reproduces both the quasi-thermal component in the keV-MeV range and the broadband non-thermal emission observed in the time-integrated and time-resolved spectra of GRB 090902B. 
A comparative analysis of GRB 240825A within a shear-acceleration dominated (strong-scattering) scenario shows that shear-accelerated electrons produce broader spectra than thermalized electrons in the weak-scattering regime. These results indicate that GRB spectral diversity likely arises from the additional emission component originating in the MJC region under different physical conditions. 
\end{abstract}

\keywords{\uat{Gamma-ray bursts}{629}; \uat{High energy astrophysics}{739}; Individuals: GRB 090902B, GRB 240825A}

\section{Introduction}\label{sec:intro}
Gamma-ray bursts (GRBs) are among the most energetic electromagnetic transients in the universe, generally associated with ultra-relativistic jets launched arising from the core collapse of massive stars or the merger of compact objects \citep{1986ApJ...308L..43P,1989Natur.340..126E,1992ApJ...395L..83N,1993ApJ...405..273W,1999ApJ...524..262M,2001ApJ...550..410M}. 
Decades of observations, particularly with the Burst And Transient Source Experiment (BATSE; 20$-$1000 keV) onboard the Compton Gamma-Ray Observatory (CGRO), have established that GRB prompt emission spectra are predominantly non-thermal and are well described by an empirical Band function, with typical peak energy of the $\nu f_\nu$ spectrum at several hundred keV \citep{1993ApJ...413..281B,2006ApJS..166..298K}. 
Broadband observations with {\em Fermi}/ GBM (Gamma-Ray Burst Monitor; 8 keV$-$40 MeV) and LAT (Large Area Telescope; 20 MeV$-$300 GeV) have identified notable deviations from this standard spectral behavior in a subset of GRBs. One primary signature of these deviations is an additional high-energy component extending into the MeV-GeV band, as observed in GRBs 080916C, 090510, 090902B, 090926A, 110731A, 130427A, 160509A, and 240825A \citep{2009Sci...323.1688A,2009ApJ...706L.138A,2010ApJ...716.1178A,2011ApJ...729..114A,2013ApJ...763...71A,2014Sci...343...42A,2018ApJ...864..163V,2025ApJ...984L..45Z}. 
Within the canonical fireball model, prompt emission is typically attributed to synchrotron ($\rm Syn$) and/or Inverse Compton ($\rm IC$) radiation from non-thermal electrons in an optically thin relativistic outflow
\citep{1994ApJ...430L..93R,1997ApJ...490...92K,2001A&A...369..694S,2002A&A...391.1141D,2011ApJ...726...90Z}. 
A quasi-thermal emission component observed in some GRBs (such as GRB 090902B and GRB 090510A) is generally thought to be photospheric emission of matter-dominated jets \citep{1986ApJ...308L..43P,1986ApJ...308L..47G,2000ApJ...530..292M,2010ApJ...709L.172R,2010ApJ...716.1178A,2011MNRAS.415.1663T,2012MNRAS.420..468P}.

The diversity of observed broadband GRB prompt spectra and the compelling evidence for a structured jet in GRB 170817A motivated our investigation of GRB emission within a structured outflow framework \citep{2017PhRvL.119p1101A,2018Natur.554..207M,2024ApJ...977..182W, 2025ApJ...981..196W}. 
Relativistic magnetohydrodynamic (MHD) and particle-in-cell (PIC) simulations demonstrate that a jet-cocoon structure self-consistently emerges as a GRB jet propagates through a material envelope \citep{2018MNRAS.477.2128H,2022MNRAS.510.4962G,2022ApJ...933L...2G,2022ApJ...933L...9G}. Turbulence, small-scale magnetic reconnection, and kinetic instabilities readily develop and persist in the jet-cocoon interaction layer \citep{2013ApJ...767...19L,2016MNRAS.456.1739B,2019MNRAS.490.4271M}. These processes facilitate effective electron acceleration, enabling them to subsequently propagate and be injected into the MJC region \citep{2013ApJ...766L..19L,2014NJPh...16c5007A,2014ApJ...783L..21S,2021ApJ...907L..44S}. 
We proposed that effective shear acceleration in the  MJC region can produce a distinct electron population. 
This population, together with electrons accelerated via the internal shocks or the magnetic process in the jet core, produces a multi-peaked non-thermal spectrum via synchrotron and synchrotron self-Compton (SSC) emission processes \citep{2024ApJ...977..182W}. Our model also well reproduces the observed bright optical emission and X-ray and GeV excesses over the Band function \citep{2025ApJ...990..157W}.

Note that efficient energization of injected electrons into a non-thermal distribution via shear acceleration in the MJC region requires a strong-scattering limit condition \citep{1989ApJ...340.1112W,2002A&A...396..833R,2004ApJ...617..155R,2018ApJ...855...31W}. In the weak-scattering limit, particle transport becomes increasingly field-aligned, with few scattering events due to long mean free paths. The acceleration efficiency fundamentally depends on the competition between the energy gain and particle escape. 
In this regime, charged particles cannot remain sufficiently confined within the shear region to repeatedly sample the velocity gradients through scattering, thereby suppressing sustained non-thermal acceleration \citep{1989ApJ...340.1112W,2006ApJ...652.1044R,2019ApJ...886L..26R}. As a result, the dissipated energy can primarily contribute to electron heating and partial thermalization, producing a quasi-thermalized electron population \citep{2003matu.book.....B,2019mlap.book.....G}. This population may contribute a distinct spectral component superimposed on the jet core emission. We focus on this scenario in this analysis. 
The paper is organized as follows. We describe the model in Section~\ref{sec:model}. As a case study, we apply the model under the weak scattering conditions to GRB 090902B in Section~\ref{sec:case}. Using GRB 240825A as an example, we compare emission spectra of shear-accelerated electrons in the strong-scattering regime with those of thermalized electrons in the weak-scattering regime in Section~\ref{sec:compare}. Conclusions and discussion are presented in Section~\ref{sec:SummaryAndDiscussion}.

\section{Emission Spectrum of thermalized electrons under weak scattering conditions in the MJC Region}\label{sec:model}
A jet-cocoon structure could be formed when the GRB ejecta propagates into the surrounding medium. Simulations reveal that, at the flow contact discontinuity, MHD instabilities, such as kink, Kelvin-Helmholtz (KH), and Rayleigh-Taylor (RT) modes, can naturally develop and persist, coexisting with sustained MHD turbulence \citep{2017PhR...720....1Z,2018PhRvL.121x5101A,2020ApJ...896L..31D}. 
These instabilities inevitably generate reconnecting current sheets and the MHD turbulence further develops small-scale sheets through dynamic alignment \citep{2017PhRvL.118x5101L,2018ApJ...854..103C,2025ARA&A..63..127S}. 
In the jet-cocoon interaction layer, electrons can be energized by magnetic reconnection and subsequently injected into the MJC region. Different from the interaction layer, the MJC region is characterized by ordered shear flows. 
The injected electrons can be further accelerated via the shear-acceleration mechanism in the strong scattering limits or further thermalized to be a quasi-thermal electron population in the MJC region.
We model the GRB outflow as a uniform jet core with opening angle $\theta_{\rm jet}$ surrounded by a cocoon with opening angle $\theta_{\rm cn}$. In the following calculations, the jet core and the MJC region are treated as independent emission components within a spherically symmetric radial structure, which consists of an ultra-relativistic jet core ($r<r_0$), a thin jet-cocoon interaction layer with strong turbulence ($r_0\lesssim r \lesssim r_1$), and a mild-relativistic MJC region ($r_1<r<r_2$).

In the weak-scattering limit, particles propagate along smooth flows with a low acceleration rate because they must traverse large distances to sample the shear\footnote{In the weak scattering limit, the shear-acceleration becomes increasingly relevant in relativistic, large-scale flows (e.g., active galactic nucleolus jet spines or galactic winds), where both the velocity difference and the available confinement volume/time are large \citep{2004ApJ...617..155R,2017ApJ...842...39L,2023ApJ...958..169W}.}. Hence, the injected electrons are not effectively accelerated by the velocity shear \citep{1989ApJ...340.1112W,2006ApJ...652.1044R}. Instead, dissipated energy is channeled into random particle motions, driving particle heating and partial thermalization\citep{2003matu.book.....B,2019mlap.book.....G}. Local small-scale turbulence and stochastic fields can promote the isotropization of electron momentum and mediate energy redistribution \citep{2013ApJ...771..124Z}. In an optically thick region, Coulomb and Compton interactions further drive electron thermalization. Under these conditions, the MJC region can maintain a radial temperature profile, which is modeled as a power-law function, e.g. 
\begin{equation}
T(r) = T_{\max}
\left( \frac{r}{r_{1}} \right)^{-q_{\rm T}}.
\end{equation}
The emission spectrum is calculated through  
\begin{equation}\
\label{Eq2}
F_{\rm th}(\nu) = \int^{T_{\max}}_{T_{\min}}\ \frac{\mathrm{d}A}{\mathrm{d}T}\, B_{\nu}(T)\, \mathrm{d}T, 
\end{equation}
where $B_\nu(T) = (2h/c^2) (\nu^3/(e^{h\nu / kT} - 1))$ is the Planck function and $A$ is the effective radiating area obtained by integrating over the emitting surface of the considered spherically symmetric MJC region.

\begin{figure}[htpb!]
        \centering
        \includegraphics[width=0.37\textwidth]{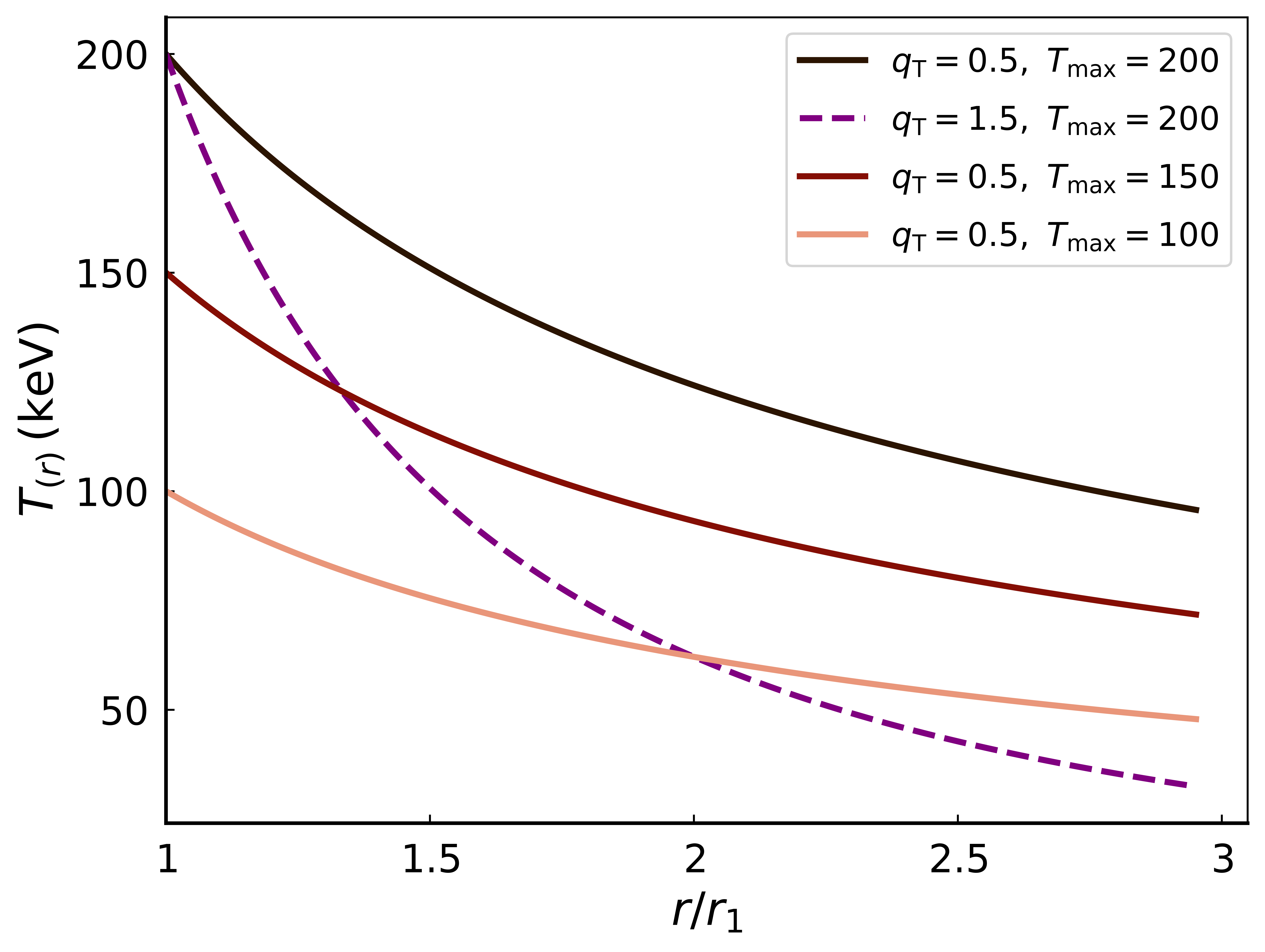}
        \includegraphics[width=0.37\textwidth]{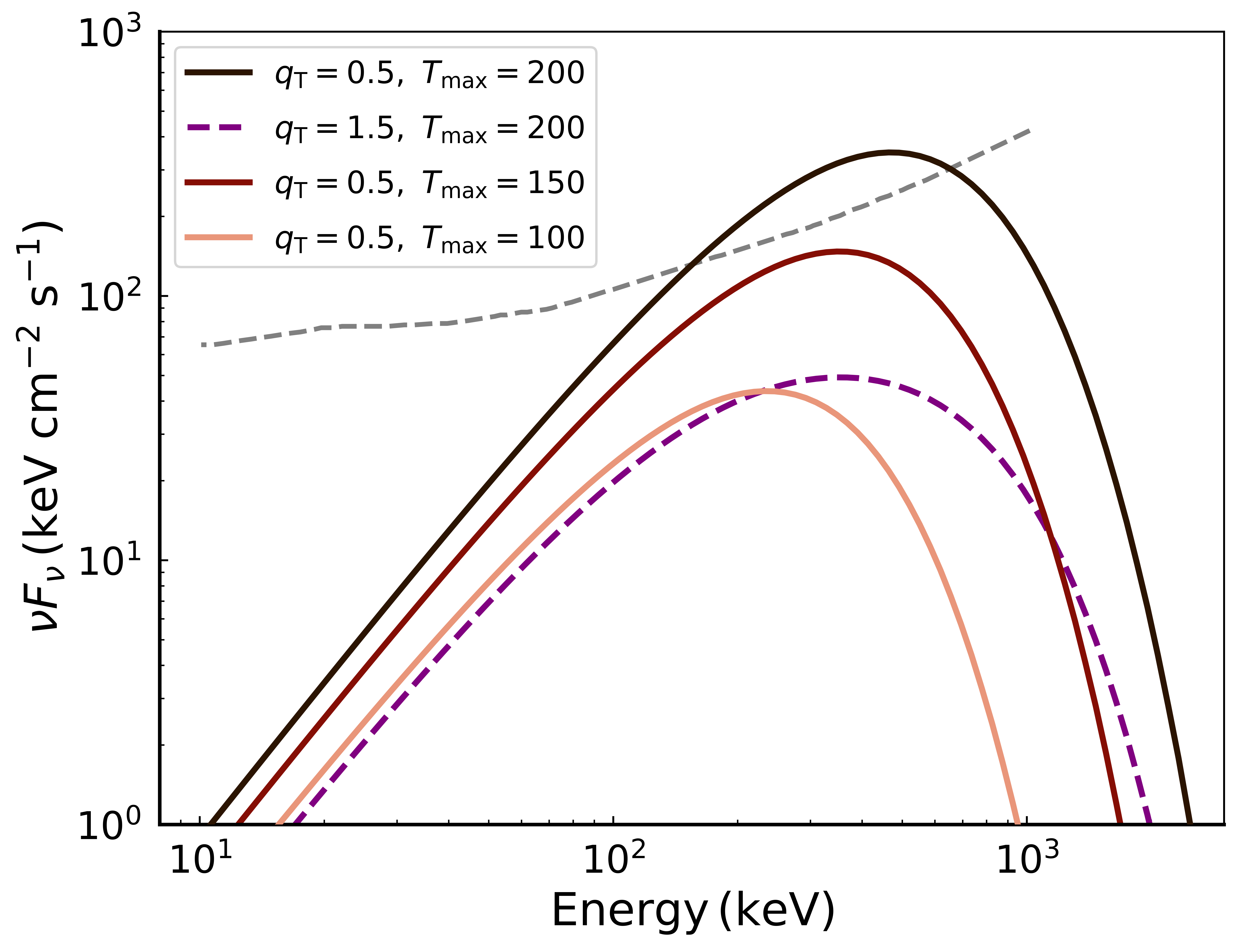} 
        \caption{Temperature profiles of electrons in the MJC region for different parameter sets ({\em Left panel}) and the corresponding emission spectra ({\em Right panel}). The gray dashed line indicates the {\em Fermi}/GBM flux threshold. }
        \label{fig:compare}
\end{figure}

Assuming the source has a redshift of $z=1$ and an emission radius of $R = 10^{15}$ cm with $\theta_{\rm jet} = 0.07$ rad and $\theta_{\rm cn} = 0.7$ rad, Figure~\ref{fig:compare} presents the temperature profiles and the corresponding emission spectra for different parameter sets. It shows that a shallower temperature profile (smaller $q_T$ value) and a higher $T_{\max}$ produce a narrower spectrum with a larger peak flux.
In the case of $\{q_T, T_{\max} \} =\{0.5, 200 {\,\rm keV}\}$, this spectral component could be robustly detectable with the {\em Fermi}/GBM. For $T_{\max}=150$ keV and $q_T=0.5$, the peak flux of this emission component falls slightly below the GBM threshold. In the other cases, the peak flux is substantially lower than the GBM threshold. 

\section{Case Study: GRB 090902B}\label{sec:case}
\begin{figure}[htpb!]
    \centering
    \begin{minipage}{0.62\textwidth}
    \centering
        \includegraphics[width=0.49\textwidth]{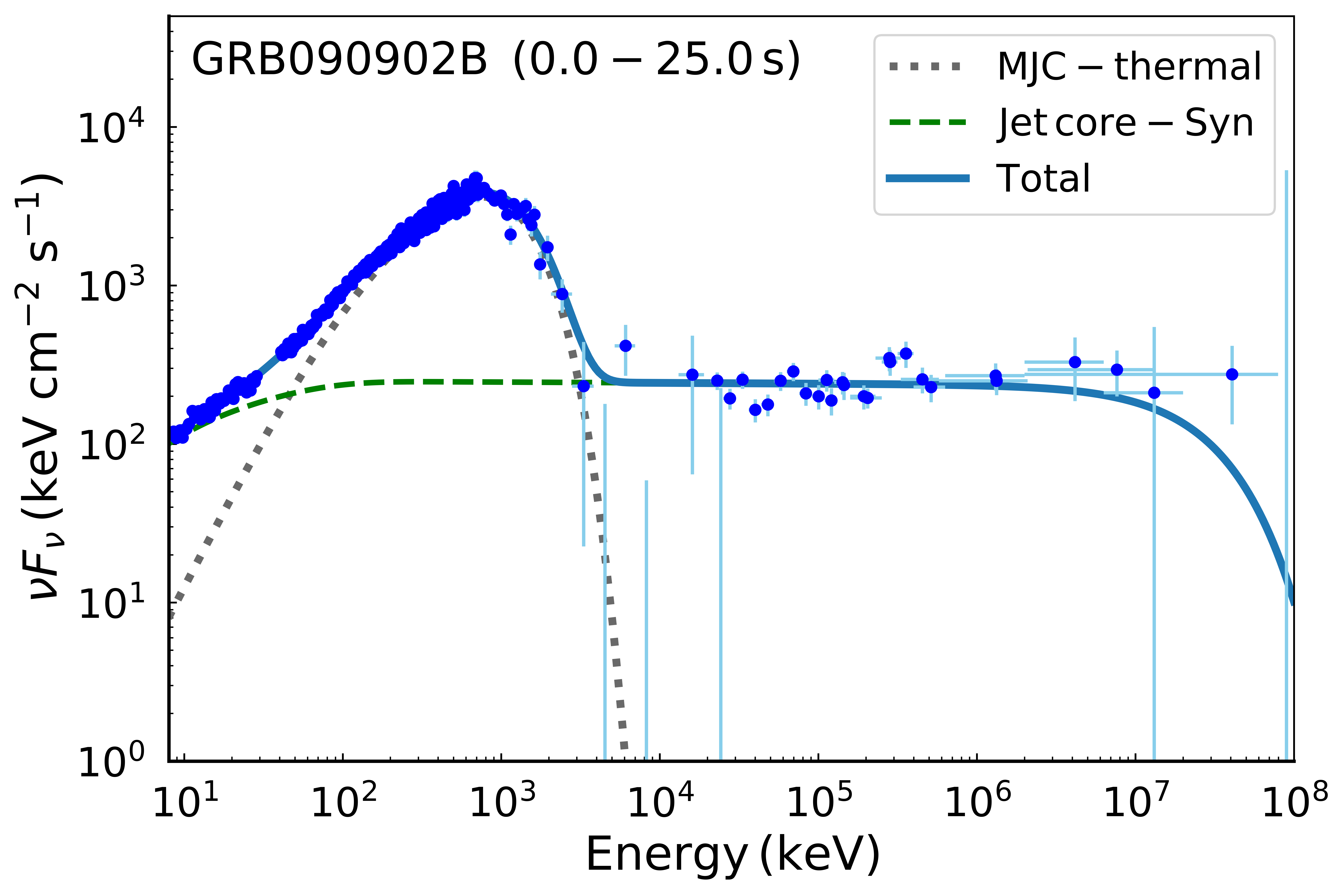}
        \includegraphics[width=0.49\textwidth]{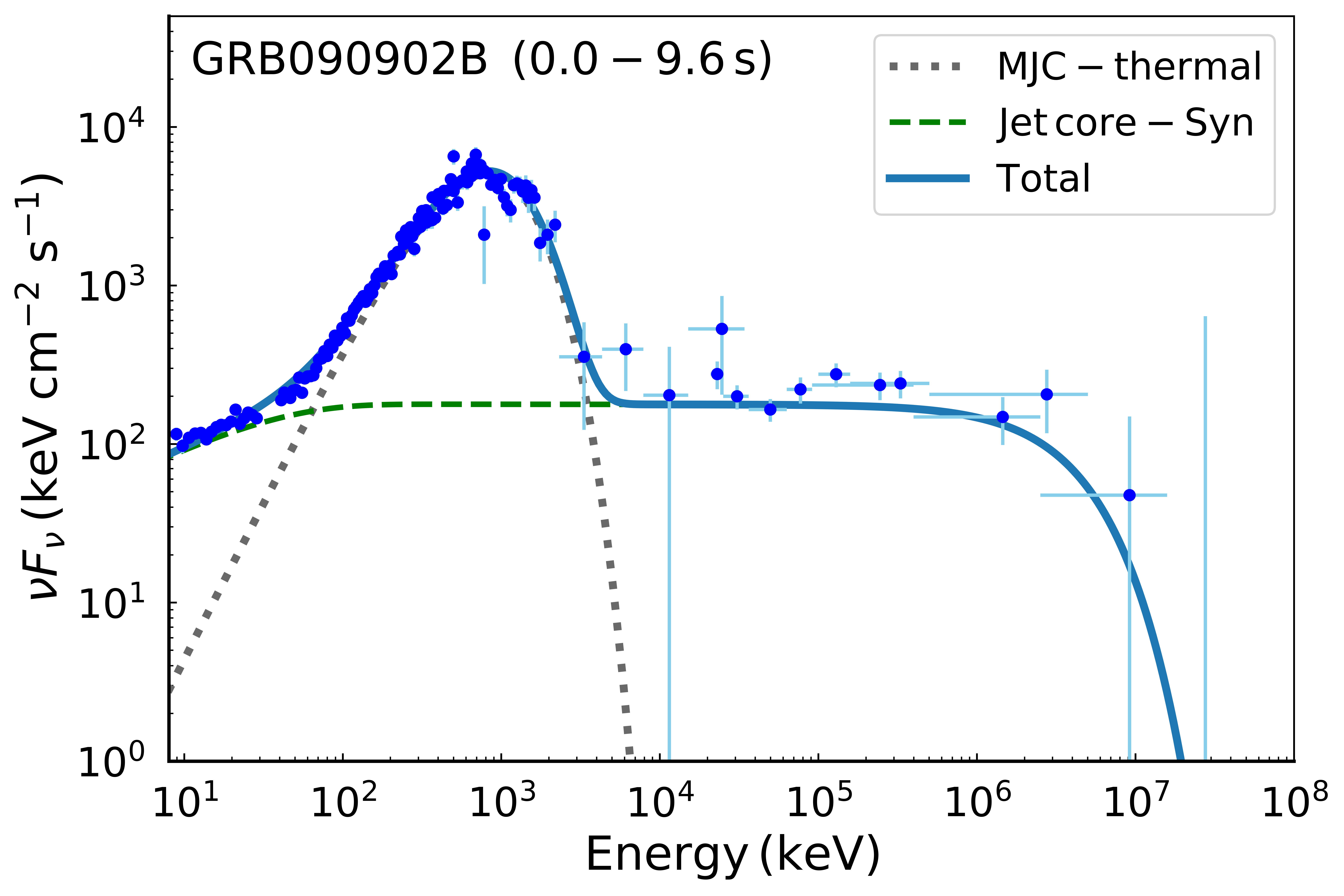}
            
        \includegraphics[width=0.49\textwidth]{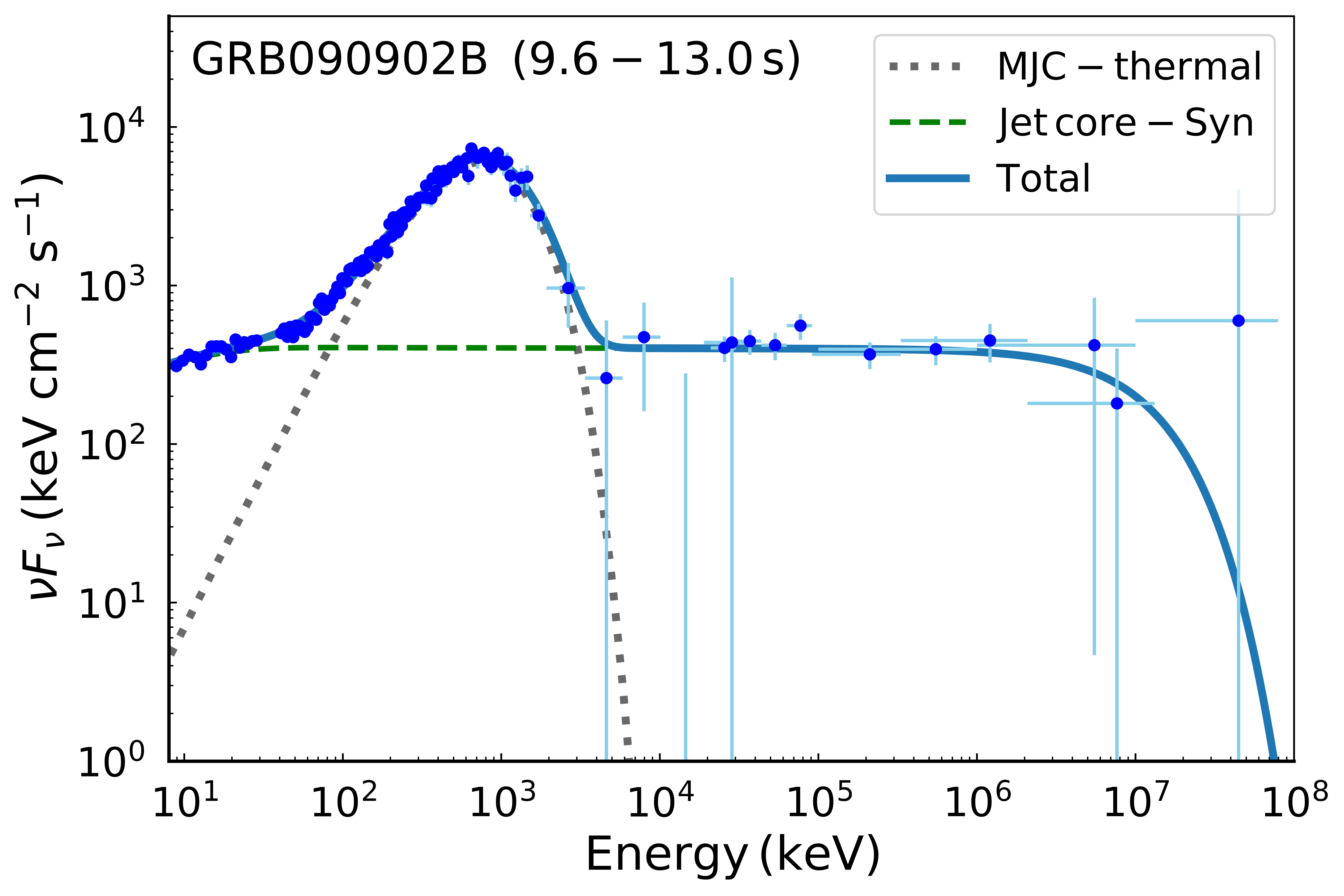}
        \includegraphics[width=0.49\textwidth]{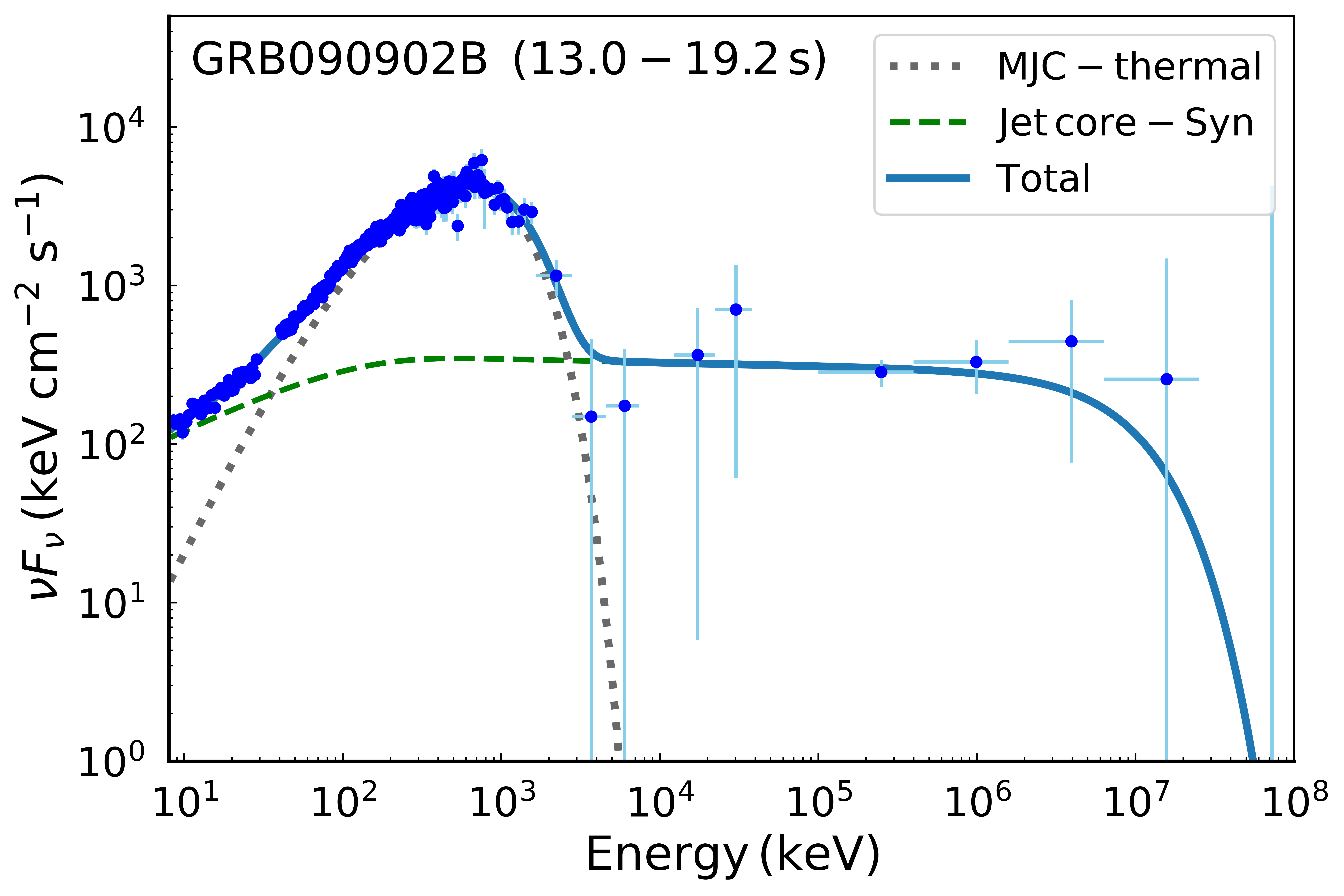}
    \end{minipage}
    \hfill
    \begin{minipage}{0.35\textwidth}
    \centering
        \includegraphics[width=\textwidth]{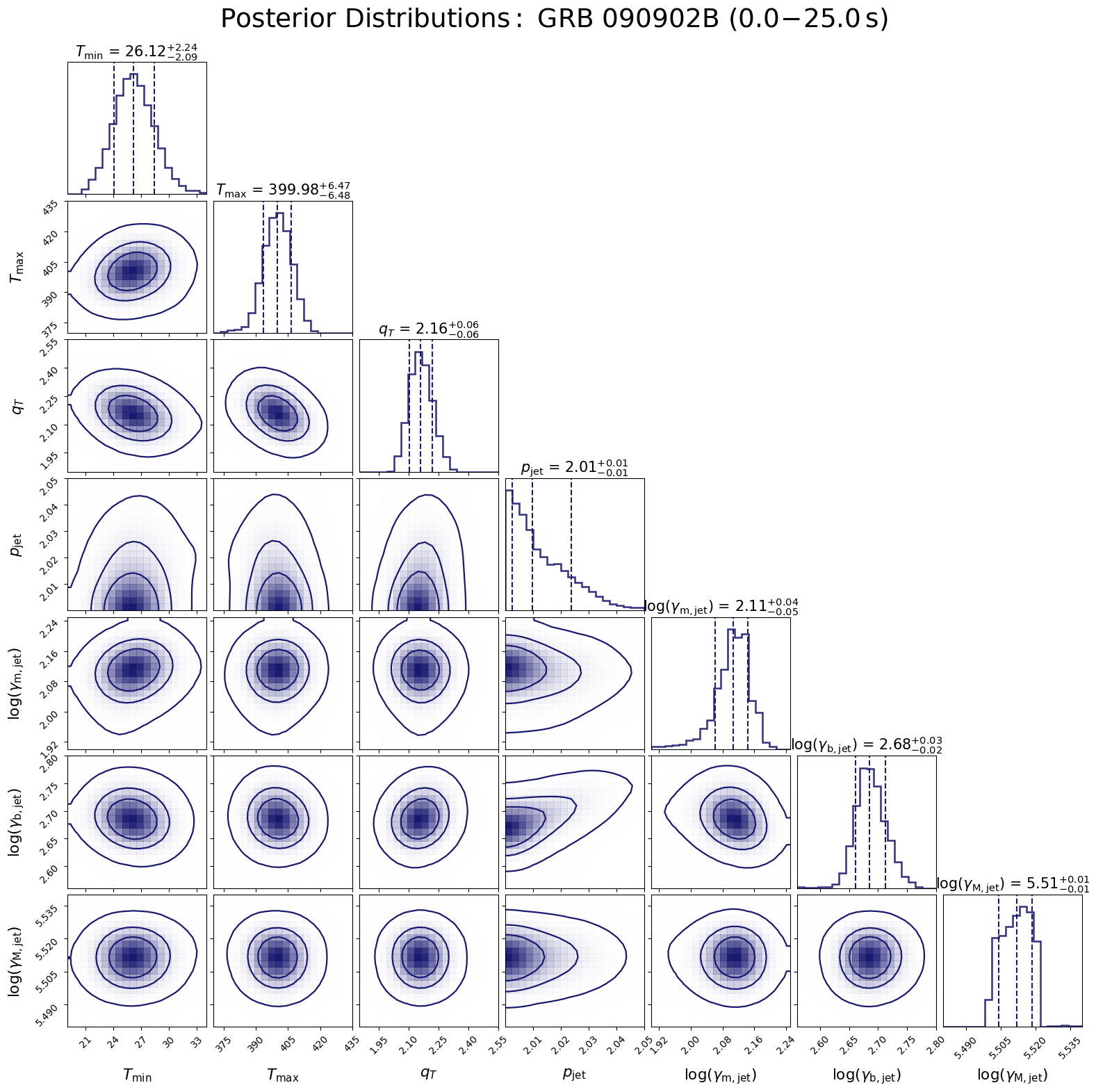}
    \end{minipage}
    \caption{Observed time-integrated and time-resolved spectra of GRB 090902B and the best-fit model spectra ({\em left and middle panels}). Dashed and dotted lines denote the jet-core and MJC emission components, and the solid line shows the total emission. The {\em right panel} illustrates the posterior distributions of the model parameters derived from the MCMC fit to the time-integrated spectrum.}
    \label{fig:090902B}
 \end{figure}

GRB 090902B is the most prominent representative event with the detection of a distinctive thermal emission component. The burst was observed by the {\em Fermi} mission. We retrieve GBM and LAT data from the Fermi GBM public data archive\footnote{\url{https://heasarc.gsfc.nasa.gov/FTP/fermi/data/gbm/daily/}} and the Fermi Science Support Center\footnote{\url{http://fermi.gsfc.nasa.gov/ssc/data/}}, respectively. GBM data reduction is performed using Gbmtools-v1.1.1 \citep{GbmDataTools}. LAT data are processed using Fermitools-v2.2.0\footnote{\url{https://fermi.gsfc.nasa.gov/ssc/data/analysis/software/}} with the P8R3\_TRANSIENT020e instrument response function. The extracted time-integrated and time-resolved spectra of GRB 090902B are shown in Figure~\ref{fig:090902B}. 

We attribute the broadband spectra of GRB 090902B to the combined contribution of synchrotron radiation of the shock-accelerated electrons in the jet core and multi-color thermal emission from the thermarized electrons in the MJC region. The emission from the jet core is modeled as synchrotron-dominated radiation in the fast-cooling scenario, with the electron distribution described by a broken power law, i.e., $dN_{e}/d\gamma_{e} \propto  
\gamma_{e,\mathrm{jet}}^{-2}$ for $ \gamma_{e,\mathrm{jet}} \leqslant \gamma_{\mathrm{b},\mathrm{jet}}$, $dN_{e}/d\gamma_{e} \propto
\gamma_{e,\mathrm{jet}}^{-p_{\mathrm{jet}}-1}$ for $\gamma_{e,\mathrm{jet}} > \gamma_{\mathrm{b},\mathrm{jet}}$, in the range of $[\gamma_{\mathrm{m},\mathrm{jet}}, \gamma_{\mathrm{M},\mathrm{jet}}]$ with a break at $\gamma_{\rm b,jet}$. 
The emission of the thermalized electrons in the MJC region is given by Eq.~\ref{Eq2}. Spectral fitting is performed via a Markov Chain Monte Carlo (MCMC) method with the \texttt{emcee} package \citep{2013PASP..125..306F}. The jet-core Lorentz factor is set to $\Gamma_{\rm jet} = 1000$ \citep{2009ApJ...706L.138A}, and the MJC region is assumed to have a bulk velocity of $\beta_{\rm cn} \sim 0.8$.  We focus on the properties of the electron population. To reduce parameter degeneracy, we set $B_{\rm jet}=6\times 10^{5}$ G and treat the remaining parameters as free \citep{2017ApJ...837...33B}. 
The derived model parameters are summarized in Table~\ref{table:090902B}, and the best-fit lightcurves are shown in Figure~\ref{fig:090902B}. Both the time-integrated and time-resolved spectra are well reproduced by the model. The emission of quasi-thermal electrons from the MJC region well accounts for the narrow, bright emission peak in the keV-MeV range. The synchrotron emission of non-thermal electrons in the jet core produces a broad component from several keV to $\sim10$ GeV. It extends down to the low-energy end of GBM and manifests as an excess over the quasi-thermal component at tens of keV.
Across the time-resolved intervals, $T_{\rm min}$ decreases significantly from 203.63 keV to 29.18 keV, while $T_{\rm max}$ declines more gradually from 454.26 keV to 367.44 keV. $q_T$ increases from 0.64 to 1.76, indicating an increased effective emitting area associated with low-temperature components. Within parameter uncertainties, no appreciable evolution is found in the electron distribution of the jet-core, while the index $p_{\rm jet}$ is poorly constrained.

\begin{table}[htbp]
        \centering
        \caption{Parameter sets derived from time-integrated and time-resolved spectral modeling of GRB 090902B. Boldface entries correspond to the time-integrated spectrum. }
        \begin{tabular}{c|c|c c c c c c c c}
        \hline\hline
        GRB & \multicolumn{1}{c|}{Times (s)} & $T_{\rm min}$(\scriptsize keV) & $T_{\rm max}$(\scriptsize keV)  & $q _{\rm T}$ & $p_{\rm jet}$ & $\mathrm{1og(}\gamma_{m, \rm jet})$ & $\mathrm{1og(}\gamma_{\rm b, jet})$ & $\mathrm{1og(}\gamma _{\rm M, jet})$ \\\hline
        
        \multirow{4}{*}{090902B} 
         & \textbf{0.0--25.0}    & $\mathbf{26.12_{-2.09}^{+2.24}}$ & $\mathbf{399.98_{-6.48}^{+6.47}}$ & $\mathbf{2.16_{-0.06}^{+0.06}}$ & $\mathbf{2.01_{-0.01}^{+0.01}}$ & $\mathbf{2.11_{-0.05}^{+0.04}}$ & $\mathbf{2.68_{-0.02}^{+0.03}}$ & $\mathbf{5.51_{-0.01}^{+0.01}}$ \\ \cline{2-9}
         & 0.0--9.6     & $203.63_{-7.42}^{+7.56}$ & $454.26_{-37.58}^{+30.84}$ & $0.64_{-0.07}^{+0.08}$   & $2.00 (\rm fixed)$ & $1.67_{-0.22}^{+0.19}$ & $2.66_{-0.04}^{+0.03}$ & $5.05_{-0.04}^{+0.10}$ \\ \cline{2-9}
         & 9.6--13.0    & $150.27_{-11.49}^{+11.00}$ & $421.01_{-34.76}^{+43.11}$ & $1.08_{-0.22}^{+0.38}$ & $2.00 (\rm fixed)$ & $1.85_{-0.29}^{+0.19}$  & $2.35_{-0.04}^{+0.04}$ & $5.32_{-0.23}^{+0.33}$ \\ \cline{2-9}
         & 13.0--19.2   & $29.18_{-2.52}^{+2.71}$  & $367.44_{-8.63}^{+9.92}$ & $1.76_{-0.07}^{+0.07}$  & $2.04_{-0.03}^{+0.03}$ & $1.79_{-0.24}^{+0.17}$ & $2.88_{-0.05}^{+0.06}$ & $5.26_{-0.14}^{+0.32}$ \\ \hline
        
        \end{tabular}
        \label{table:090902B}
\end{table}

\section{Comparison of emission spectra of thermalized and shear-accelerated electrons}\label{sec:compare}
As described above, in the weak-scattering regime, the quasi-thermal electron population in the MJC region can account for the narrow keV-MeV spectral component of GRB 090902B. As shown in \cite{2024ApJ...977..182W}, in the strong-scattering regime, shear-accelerated electrons in the MJC region can produce a distinct low-energy non-thermal hump beyond the jet-core emission, shaping a bimodal prompt gamma-ray spectrum in some GRBs. GRB 240825A is a representative example with such a bimodal feature. We further present a comparative model analysis of this burst with GRB 090902B.

\begin{figure}[htpb!]
    \centering
    \begin{minipage}{0.62\textwidth}
    \centering
        \includegraphics[width=0.49\textwidth]{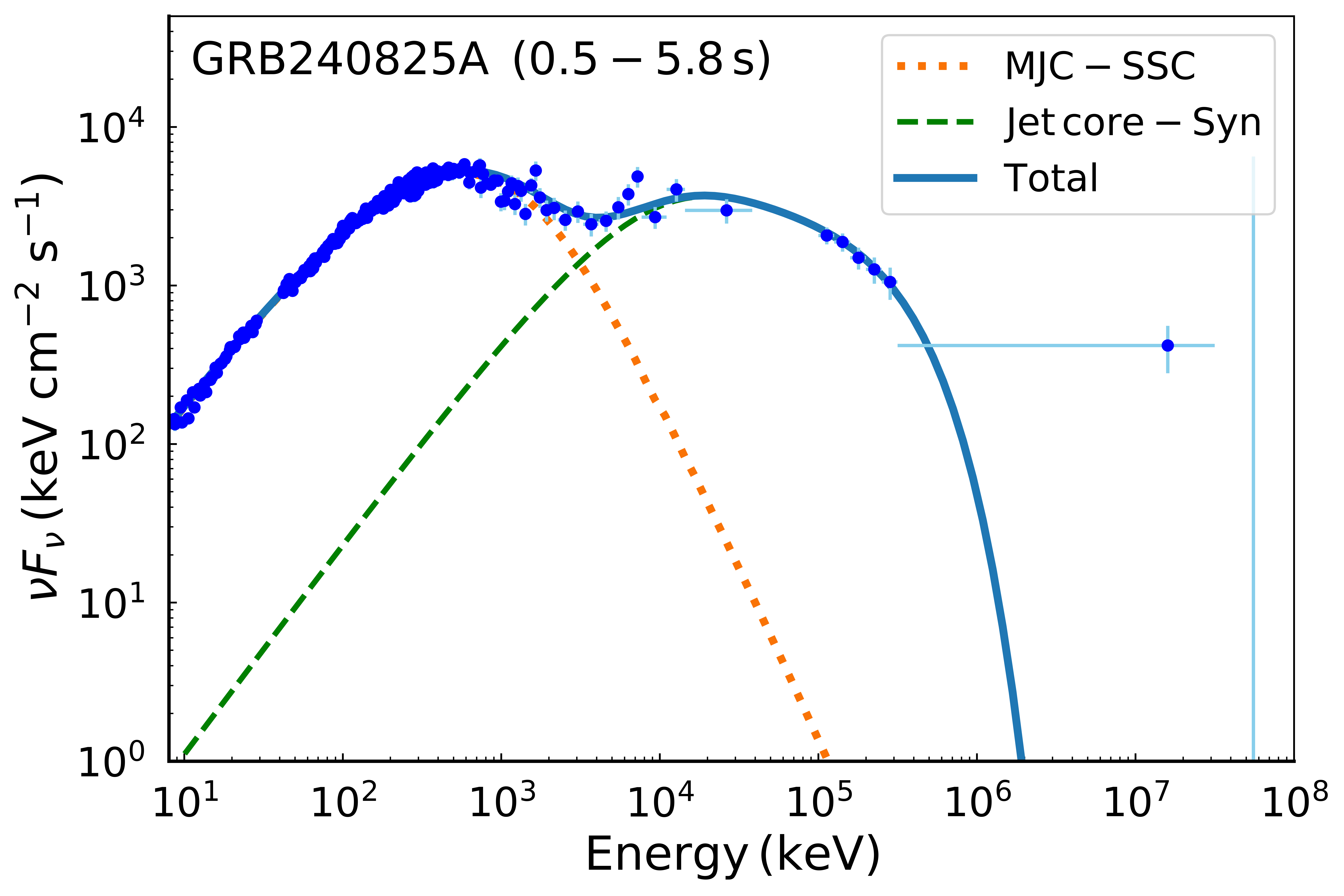}
        \includegraphics[width=0.49\textwidth]{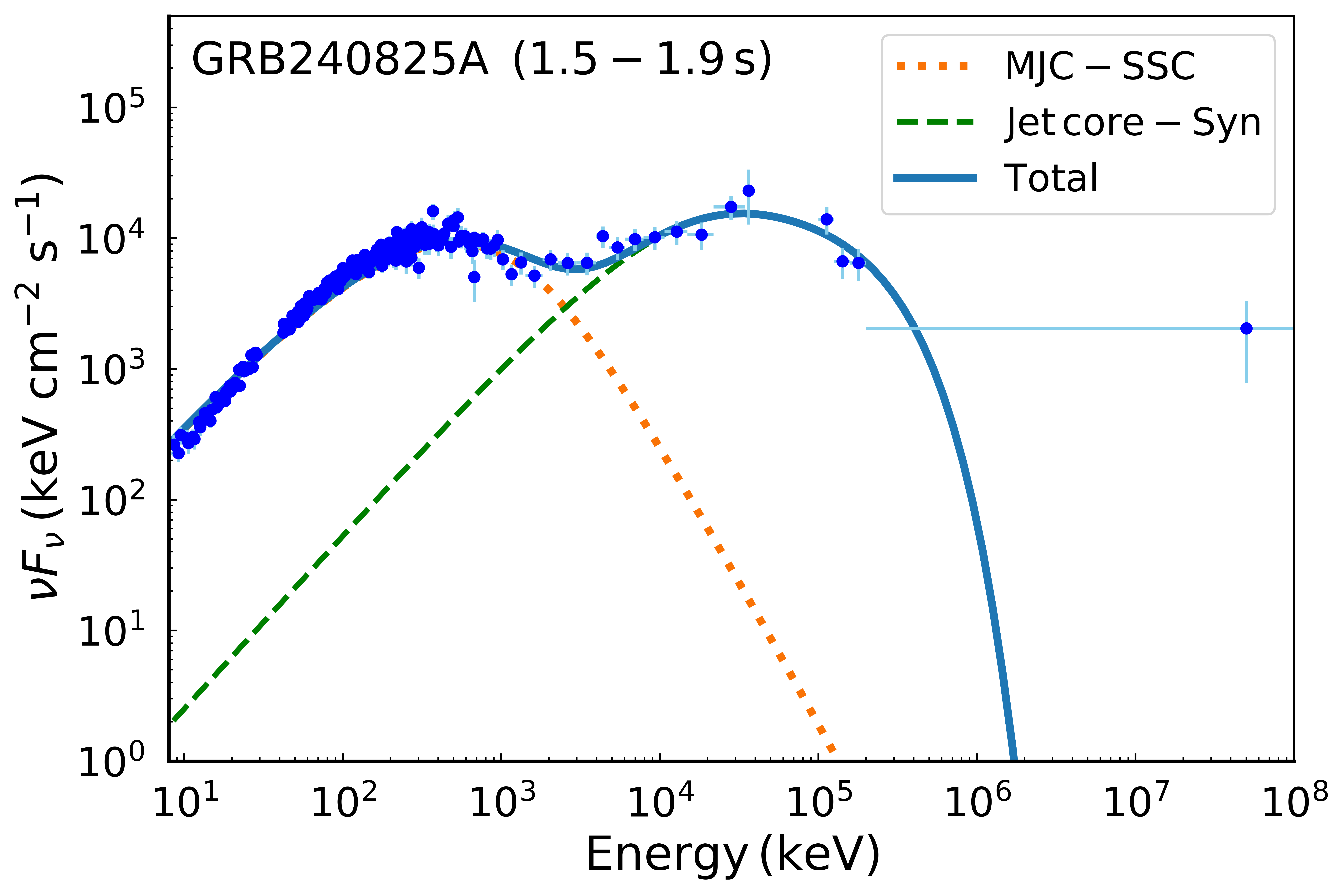}
            
        \includegraphics[width=0.49\textwidth]{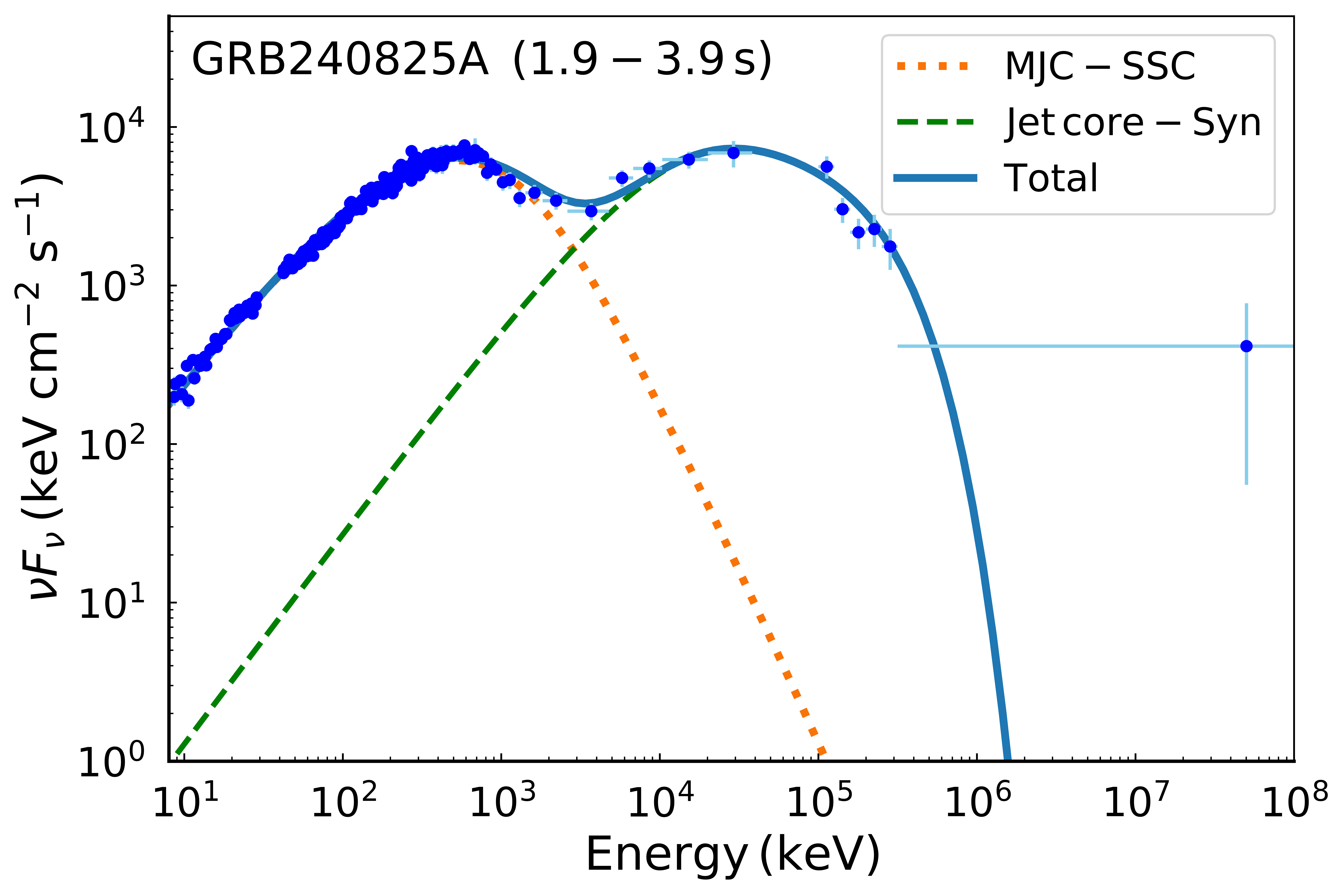} 
        \includegraphics[width=0.49\textwidth]{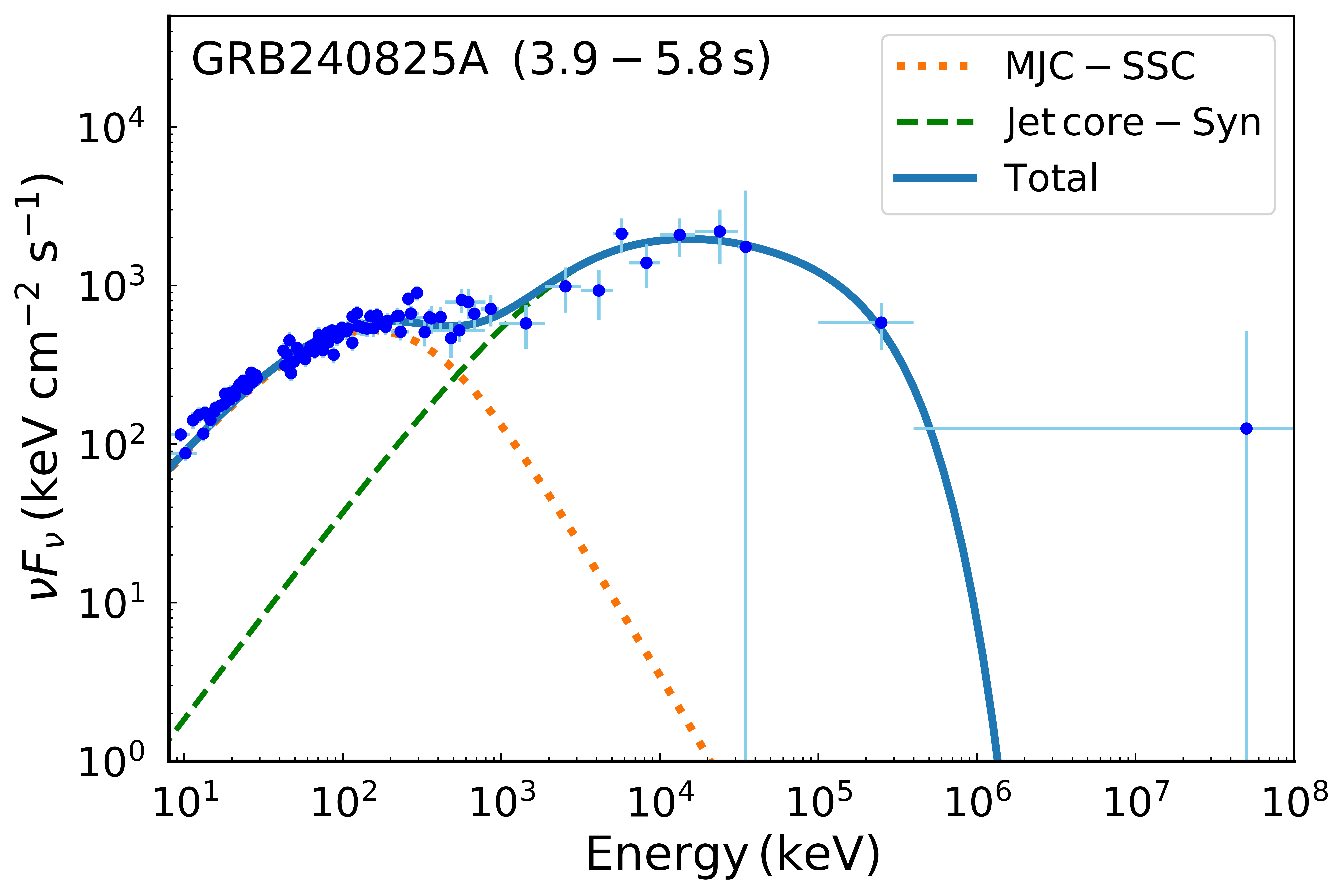}\\
    \end{minipage}
    \hfill
    \begin{minipage}{0.35\textwidth}
    \centering
        \includegraphics[width=\textwidth]{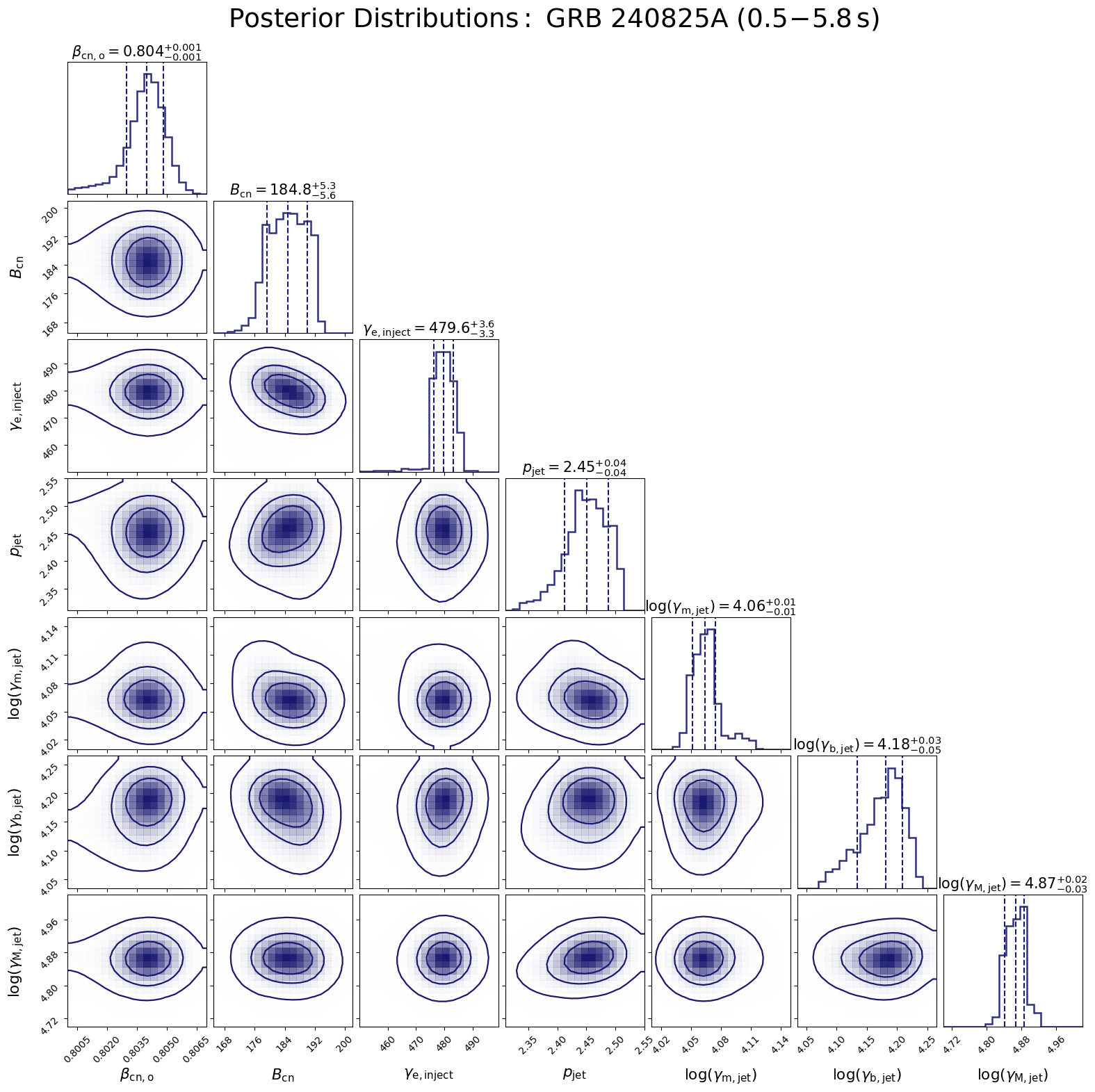}
    \end{minipage}

    \caption{Similar to Figure~\ref{fig:090902B} but for GRB 240825A. }
    \label{fig:240825A}
\end{figure}

Similarly, we perform time-integrated and time-resolved spectral modeling of GRB 240825A within the strong-scattering scenario. Following \cite{2024ApJ...977..182W}, the radial velocity profile is modeled as an exponential function
$u_{\rm cn}(r)=\beta _{\rm cn,0}e^{-k}$, where $k = (r/r_2) \ln (\beta_{\rm cn,0}/\beta_{\rm cn,2})$, $\beta _{\rm cn,0}$ and $\beta _{\rm cn,2}$ are the fluid velocities at $r_{0}$ and $r_{2}$. 
We adopt $\Gamma_{\rm jet} = 500$ as a reference value and set $B_{\rm jet}=2.6\times 10^{6}$ G in the MCMC fitting \citep{2018A&A...609A.112G}. 
Figure~\ref{fig:240825A} shows the time-integrated and time-resolved spectra of GRB 240825A with the best-fit model curves. The derived parameters are listed in Table~\ref{table:240825A}.
Both the time-integrated and time-resolved spectra can be well represented. The low-energy hump in the keV-MeV band is attributed to the SSC emission of the shear-accelerated electrons in the MJC region, and the high-energy component above MeV arises from shock-accelerated electrons in the jet core.

\begin{figure}[htpb!]
        \centering
        \includegraphics[width=0.43\textwidth]{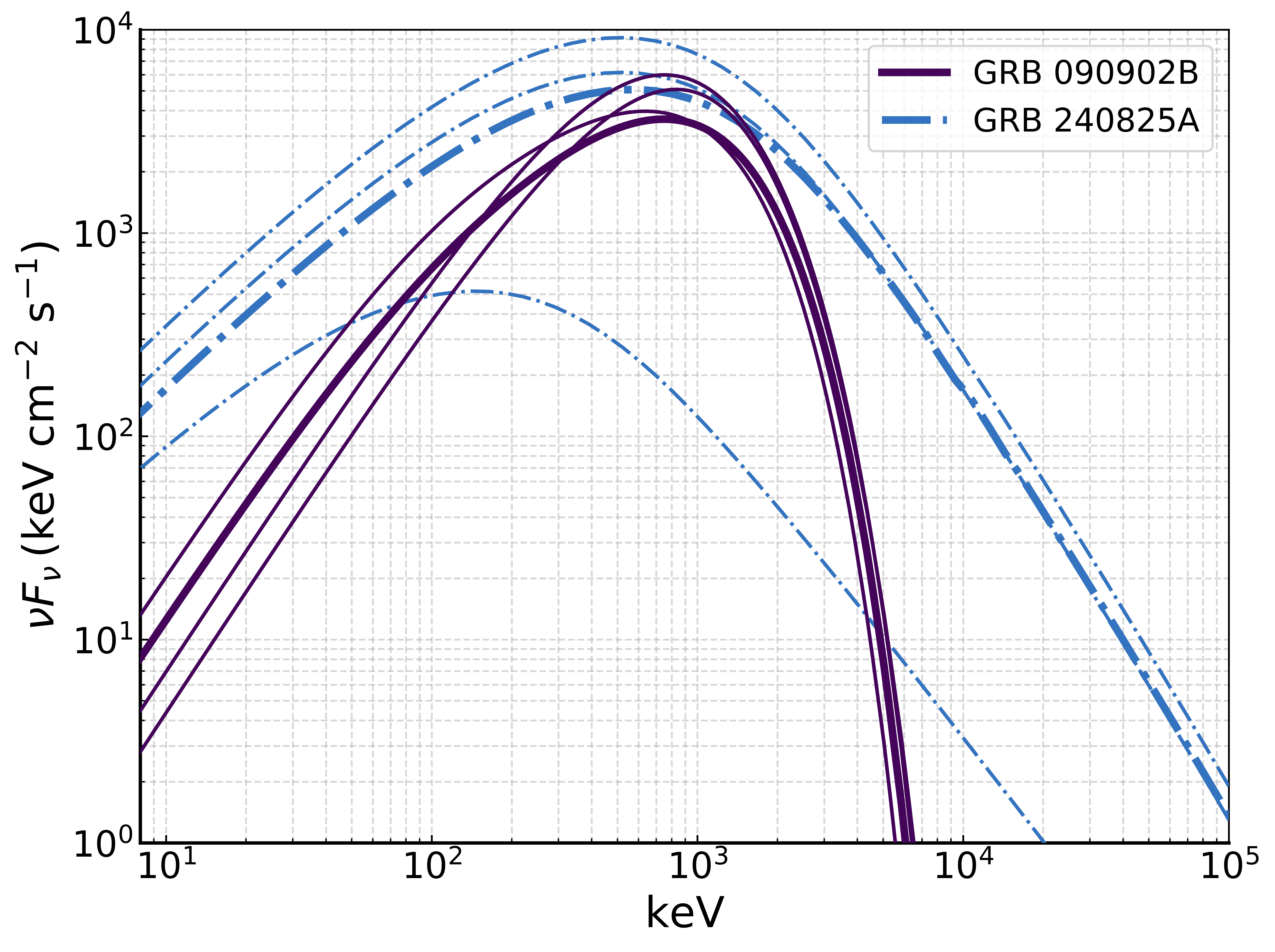}
        \caption{ Spectral profiles of emission in the MJC region of GRB 090902B (thermalized dominated) and GRB 240825A (shear-accelerated dominated). Thick and thin lines denote time-integrated and time-resolved spectra, respectively. }
        \label{fig:compare_model}
\end{figure}

Figure~\ref{fig:compare_model} compares the emission spectra of the MJC region extracted from our modeling for GRBs 090902B and 240825A. The emission spectrum of shear-accelerated electrons in GRB 240825A is considerably broader than that from thermalized electrons in GRB 090902B.

\begin{table}[htbp]
        \centering
        \caption{Parameter sets derived from time-integrated and time-resolved spectral modeling of GRB 240825A. Boldface entries correspond to the time-integrated spectrum.}
        \begin{tabular}{c|c|c c c c c c c c}
        \hline\hline
        GRB & \multicolumn{1}{c|}{Times (s)} & $\beta _{\mathrm{cn},0}$ & $B_{\mathrm{cn}}$(\scriptsize G) & $\gamma _{e,\mathrm{i nject}}$ & $p_{\rm jet}$ & $\mathrm{1og(}\gamma_{m, \rm jet})$ & $\mathrm{1og(}\gamma_{\rm b, jet})$ & $\mathrm{1og(}\gamma _{\rm M, jet})$ \\
        \hline
         
        \multirow{4}{*}{240825A} 
         & \textbf{0.5--5.8}     & $\mathbf{0.804_{-0.001}^{+0.001}}$ & $\mathbf{184.8_{-5.6}^{+5.3}}$ & $\mathbf{479.6_{-3.3}^{+3.6}}$  & $\mathbf{2.45_{-0.04}^{+0.04}}$ & $\mathbf{4.06_{-0.01}^{+0.01}}$  & $\mathbf{4.18_{-0.05}^{+0.03}}$ & $\mathbf{4.87_{-0.03}^{+0.02}}$ \\ \cline{2-9}
         & 1.5--1.9   & $0.803_{-0.002}^{+0.001}$ & $140.3_{-3.2}^{+21.8}$ & $500.9_{-22.5}^{+1.7}$  & $2.21_{-0.08}^{+0.04}$ & $4.12_{-0.01}^{+0.03}$  & $4.34_{-0.02}^{+0.04}$ & $4.79_{-0.02}^{+0.05}$ \\ \cline{2-9}
         & 1.9--3.9   & $0.802_{-0.002}^{+0.002}$ & $169.9_{-15.0}^{+4.0}$ & $478.0_{-10.9}^{+5.5}$ & $2.31_{-0.02}^{+0.03}$ & $4.14_{-0.01}^{+0.02}$  & $4.29_{-0.03}^{+0.02}$ & $4.80_{-0.02}^{+0.03}$ \\ \cline{2-9}
         & 3.9--5.8   & $0.822_{-0.002}^{+0.005}$ & $63.1_{-1.5}^{+3.0}$  & $441.9_{-3.0}^{+2.1}$    & $2.25_{-0.12}^{+0.05}$ & $3.79_{-0.08}^{+0.05}$  & $4.19_{-0.05}^{+0.03}$ & $4.81_{-0.04}^{+0.03}$ \\ \hline

        \end{tabular}
        \label{table:240825A}
\end{table}

\section{Conclusions and Discussion} \label{sec:SummaryAndDiscussion}
Relativistic simulations indicate that a jet-cocoon structure self-consistently forms as a GRB jet propagates through a material envelope. Considering that the GRB ejecta consists of an ultra-relativistic uniform core surrounded by a structured cocoon, we investigate the emission properties of the thermalized electrons in the MJC region under the weak-scattering regime. 
The radial temperature profile of the MJC electron population is characterized as a power-law function, $T(r) = T_{\max}(r/r_{1})^{-q_{\rm T}}$, in the range $[T_{\min}, T_{\max}]$. The effective radiating area is modeled as the integral over the emitting surface of a spherically symmetric MJC structure. We demonstrate that both the spectral width and the peak flux of the emission spectrum are sensitive to the parameters $q_T$ and $T_{\max}$. 

Incorporating the thermal emission of the thermlized electron in the MJC region and the synchrotron emission from the shock-accelerated electrons in the jet core, we interpret the keV-MeV-GeV spectrum of GRB 090902B, a typical burst featuring a bright quasi-thermal component in its prompt emission. Both the time-integrated and time-resolved spectra are well reproduced through MCMC fitting of our model, with the quasi-thermal emission in the keV-MeV band attributed to thermalized electrons in the MJC region. Across the analyzed time intervals, $T_{\rm min}$ decreases from 203.63 keV to 29.18 keV, $T_{\rm max}$ also shows a decreasing trend, from 454.26 keV to 367.44 keV, while $q_T$ increases from 0.64 to 1.76. The synchrotron emission of non-thermal electrons in the jet core produces a broad component from several keV to $\sim10$ GeV. It extends down to the low-energy end of GBM and manifests as an excess over the quasi-thermal component below $\sim 30$ keV. No appreciable evolution is observed in the jet-core electron distribution. A comparative analysis of the non-thermal emission of shear-accelerated electrons in the MJC region under the strong-scattering scenario is presented using GRB 240825A, whose prompt gamma-ray spectrum is bimodal. The result shows that shear-accelerated electrons produce broader spectra than thermalized electrons in the weak-scattering regime.

These results indicate that GRB spectral diversity likely arises from the additional emission component of electrons in the MJC region. This electron population can assume the form of a quasi-thermal population under the weak-scattering limit ($\Omega \tau_s \ll 1$) or as a non-thermal population in the strong-scattering limit ($\Omega \tau_s \gg 1$), where $\Omega$ is the gyrofrequency and $\tau_s$ is the mean scattering time. Both $\Omega$  and $\tau_s$ depend on the magnetic-field properties and the local turbulence level in the MJC region \citep{1989ApJ...340.1112W,2019Galax...7...78R}. The strong-scattering limit tends to occur in moderate baryon loaded plasma, where sufficient magnetic energy density can sustain magnetic fluctuations across spatial scales \citep{2012ApJ...744...32Z,2012SSRv..173..557L}. These fluctuations can grow through MHD instabilities, promoting the development of turbulence. Under such conditions, efficient pitch-angle scattering arises, with particles undergoing multiple scatterings and crossing regions with different bulk velocities, which can systematically increase particle energy \citep{2004ApJ...617..155R}. 
The velocity shear profile further modulates the particle acceleration. Strong shear with a large velocity gradient increases electron energizing efficiency and facilitates kinetic instabilities that amplify small-scale magnetic turbulence, allowing sustained non-thermal acceleration \citep{2004ApJ...617..155R,2009ApJ...698..293V}. 
In contrast, the weak-scattering limit typically satisfies in regions with weak magnetic turbulence and suppressed magnetic fluctuations, involving a high-baryon-loading environment.
Under such conditions, the large plasma inertia and relatively low magnetic energy density can, to some extent, impede the growth of small-scale magnetic fluctuations and suppress turbulence development \citep{2003MNRAS.345..325C,2013SSRv..178..163B,2019Galax...7...78R}. Energy dissipation within the shear flow operates in a regime of inefficient electron energy redistribution, resulting in a weakly anisotropic, quasi-thermal electron distribution. The high matter density favors optically thick conditions, where repeated Compton scattering between quasi-thermal electrons and photons further broadens the electron spectrum\citep{1979rpa..book.....R}.

Our analysis suggests that GRB broadband spectra result from the combined emission of synchrotron and SSC radiation from shock-accelerated electrons in the ultra-relativistic jet core, and either synchrotron and SSC radiation from shear-accelerated electrons or multi-color thermal emission from thermalized electrons in the mild-relativistic MJC region. This analysis, together with previous studies in this series
\citep{2024ApJ...977..182W, 2025ApJ...981..196W,2025ApJ...990..157W}, indicates that GRB spectral diversity is shaped by a combination of factors, including the jet structure and composition, the surrounding environment, and the observer viewing angle. The keV-MeV spectra of a significant fraction of GRBs are well described by the empirical Band function \citep{1993ApJ...413..281B,2006ApJS..166..298K}, yet the derived low-energy spectral index often exceeds the ``death line" of the standard synchrotron model. Attributing the keV-MeV emission to synchrotron radiation of jet-core electrons, the additional MJC emission naturally explains this deviation. 
In the conventional ``hot" fireball model, a distinct thermal component is typically interpreted as photospheric emission from matter-dominated jets \citep{1986ApJ...308L..43P,1986ApJ...308L..47G,2000ApJ...530..292M,2002ApJ...578..812M,2011MNRAS.415.1663T}. In contrast, our model attributes this component to thermalized electrons in the MJC region.

The quasi-thermal component can serve as a diagnostic probe of outflow microphysics, particularly the scattering condition and acceleration efficiency. The temporal evolution of the thermalized-electron temperature profile also reflects the dynamical evolution of the MJC region \footnote{The present model adopts radially independent, spherically symmetric components. In realistic scenarios, the jet-cocoon system may feature complex structures, such as Gaussian or power-law profiles, among others, requiring advanced numerical simulations for detailed characterization. Nevertheless, the qualitative insights of the investigation are expected to remain unchanged.}, including pressure-driven expansion and energy dissipation. 
These effects collectively shape the radial electron temperature structure and effective radiating surface as illustrated in Figure~\ref{fig:compare}. 
We note that thermal/quasi-thermal emission is rarely detected in the current GRB sample, with GRB 090902B representing the clearest case. Weak signatures of this emission have also been reported in some GRBs (e.g., \citealt{2009ApJ...702.1211R,2011ApJ...727L..33G,2014ApJ...795..155P,2018ApJ...866...13H}). This suggests that most GRBs exhibit either steep radial temperature profiles (large $q_T$) or relatively low characteristic temperatures (small $T_{\max}$), rendering this component difficult to detect with current instruments.
Instead, in the strong scattering scenario, shear-accelerated electrons can produce additional non-thermal emission beyond the jet core, potentially explaining X-ray/GeV excesses and deviations from synchrotron limits.
The combination of emission from both the jet core and MJC region, together with difference in microphysics, provide a framework to interpret GRB spectral diversity.

\begin{acknowledgements}
This work is supported by the National Key R\&D Program (2024YFA1611704 and 2024YFA1611700), the National Natural Science Foundation of China (grant Nos. 12133003 and 12203015), the Guangxi Science and Technology Innovation Platform Program (Leitai Action Plan, No. Guike LT2600640026), Guangxi Key R\&D Program (Guike FN2504240040), and the ``Guangxi Highland of Innovation Talents'' Program.
\end{acknowledgements}

\software{Astropy \citep{2013A&A...558A..33A,2018AJ....156..123A,2022ApJ...935..167A},  
          Numpy \citep{2011CSE....13b..22V}, 
          Scipy \citep{2020NatMe..17..261V}
          }

\clearpage
\bibliography{sample7}{}
\bibliographystyle{aasjournalv7}


\end{document}